\documentclass[%
 reprint,
 amsmath,amssymb,
 aps,
]{revtex4-2}

\usepackage{graphicx}
\usepackage{dcolumn}
\usepackage{bm}

\usepackage{verbatim}
\usepackage{algorithm}
\usepackage{algorithmic}
\usepackage{xcolor}

\begin{document}

\preprint{APS/123-QED}

\title{Do surface gravity waves have a frozen turbulence state?}

\author{Zhou Zhang}
\author{Yulin Pan}%
 \email{yulinpan@umich.edu}
\affiliation{%
 Department of Naval Architecture and Marine Engineering, University of Michigan
}%





\begin{abstract}
We study the energy transfer by exact resonances for surface gravity waves in a finite periodic spatial domain. Based on a kinematic model simulating the generation of active wave modes in a finite discrete wavenumber space $\mathcal{S}_R$, we examine the possibility of direct and inverse energy cascades. More specifically, we set an initially excited region which iteratively spreads energy to wave modes in $\mathcal{S}_R$ through exact resonances. At each iteration, we first activate new modes from scale resonances (which generate modes with new lengths), then consider two bounding situations for angle resonances (which transfer energy at the same length scale): the lower bound where no angle resonance is included and the upper bound where all modes with the same length as any active mode are excited. Such a strategy is essential to enable the computation for a large domain $\mathcal{S}_R$ with the maximum wavenumber $R\sim 10^3$. We show that for both direct and inverse cascades, the energy cascade to the boundaries of $\mathcal{S}_R$ can be established when the initially excited region is sufficiently large, otherwise a frozen turbulence state occurs, with a sharp transition between the two regimes especially for the direct cascade. Through a study on the structure of resonant quartets, the mechanism associated with the sharp transition and the role of angular energy transfer in the cascades are elucidated.

\end{abstract}

\maketitle


\section{\label{sec:intro}Introduction}

Wave turbulence theory (WTT) describes the statistical behavior of a large number of dispersive waves subject to nonlinear interactions \cite{zakharov2012kolmogorov}. In the framework of WTT, the wave system is described by a kinetic equation (KE) which has stationary solutions of the Kolmogorov-Zakharov power-law spectra in the inertial range. Analogous to the Kolmogorov description of hydrodynamic turbulence, WTT predicts energy cascades through scales with a constant energy flux in the inertial range. This methodology has been applied to a wide variety of physical systems including surface gravity waves \cite{zakharov1968stability}, capillary waves \cite{zakharov1967weak}, internal gravity waves \cite{lvov2004energy}, plasma waves \cite{galtier2000weak} and gravitational waves \cite{galtier2017turbulence}.

In spite of its fruitful applications, WTT is based on the assumption of an infinite spatial domain which is usually not achieved in both numerical and laboratory experiments. For a finite system, the wavenumber space becomes discrete with the spacing between adjacent wave modes $\Delta k$ determined from the size of the domain. In the weak nonlinearity limit, the energy cascade can only be excited through exact resonances, i.e., modal interactions satisfying
\begin{equation}
\begin{aligned}
    \bm{k}_1\pm\bm{k}_2\pm...\pm\bm{k}_s&=0, \\
    \omega_1\pm\omega_2\pm...\pm\omega_s&=0,
\end{aligned}
\label{eq:resonance}
\end{equation}
where $\bm{k}_i=(m_i\Delta k, n_i\Delta k)$ is the discrete wavenumber vector with $m_i,n_i\in\mathbb{Z}$ and $\omega_i$ is the angular frequency determined from the dispersion relation, say $\omega_i\sim k_i^\alpha$ with $k_i=|\bm{k}_i|$. The importance of \eqref{eq:resonance} in understanding the wave dynamics at low nonlinearity has been demonstrated in capillary wave turbulence \cite{pushkarev2000turbulence,pan2014direct,pan2015decaying}, MMT turbulence \cite{hrabski2020effect} and surface gravity wave turbulence \cite{denissenko2007gravity,hassaini2018confinement,zhang2022numerical} (also see review in \cite{falcon2022experiments}).

The study on the exact resonances in discrete wavenumber space is a main subject of discrete wave turbulence (DWT), introduced first by Kartashova \cite{kartashova1991properties,kartashova1994weakly,kartashova1998wave}. In general, the problem of computing exact resonances in a discrete wavenumber space is to find integer solutions of \eqref{eq:resonance}, which is essentially a system of Diophantine equations. The properties of solutions in different wave systems are determined by the values of $s$ in \eqref{eq:resonance} and $\alpha$ in the dispersion relation. In particular, there are certain wave systems in which no exact resonance exists (e.g., capillary waves with $s=3$, $\alpha=3/2$) \cite{kartashova1998wave}, as well as systems containing a large number of exact resonances so that the energy can be transferred to arbitrarily large wavenumbers (e.g., nonlinear Schr\"odinger (NLS) equation with $s=4$, $\alpha=2$) \cite{carles2012energy,colliander2010transfer}.

The situation for surface gravity waves ($s=4$, $\alpha=1/2$) is much more complicated. While it is known that sparse exact resonances exist for gravity waves, it is not clear what should be expected from these exact resonances in a discrete wavenumber domain $\mathcal{S}_R$ (given some initially active modes), i.e., whether a state of unlimited energy cascade or frozen turbulence is relevant. More precisely, here we define an energy transfer to reach the maximum wavenumber $R$ in $\mathcal{S}_R$ as "unlimited energy cascade (in $\mathcal{S}_R$)" and otherwise as "frozen turbulence" (which is a generalized definition relative to \cite{pushkarev2000turbulence} for capillary waves). For $R\rightarrow \infty$, this is a very hard number theoretic problem (through private communication with number theorists) which is beyond our scope in this paper. We will instead consider an $\mathcal{S}_R$ with finite $R\sim 10^3$ that is compatible with (or beyond) most state-of-the-art simulations \cite{onorato2002freely,dyachenko2004weak,yokoyama2004statistics,lvov2006discreteness,zhang2022numerical,hrabski2022properties,nazarenko2014bose,galtier2021direct,tsubota2017numerical} performed for wave turbulence.

An effective approach to study the above problem numerically is the so-called kinematic method, i.e., with an initial set of active modes, the energy spreading to outer regions is computed iteratively by finding solutions to \eqref{eq:resonance}. This method has been applied to capillary waves \cite{connaughton2001discreteness}, gravity waves \cite{lvov2006discreteness} and NLS \cite{hrabski2021energy}, which however mainly focus on the effect of exact/quasi-resonances in a small domain, i.e., with small $R$. As an example, for gravity waves, the only available result is for $R=64$ with initially active modes in $k\in [6,9]$, where it is shown in \cite{lvov2006discreteness} that a frozen turbulence state is observed. The difficulty in extending the existing kinematic method to our interest of $R\sim 10^3$ lies in the computational cost. A brute-force search of solution of \eqref{eq:resonance}, as applied in \cite{lvov2006discreteness}, exhibits a computational cost that grows fast with the increase of the number of active modes (in particular the cost scales as the cubic of active mode numbers for each iteration). Therefore, in order to make cases of $R\sim 10^3$ computationally tractable, a much more efficient numerical method needs to be developed for the kinematic study.

One way to save some computational cost is to pre-compute the set $Q$ of all exact resonances within $\mathcal{S}_R$ using a fast generic method \cite{kartashova2006laminated,kartashova2007laminated2,kartashova2007laminated3}. With the invariant set $Q$ available, we can loop over $Q$ to compute the energy spreading in each iteration, instead of a much larger set of all combinations of the activated modes (that expands up to $O(10^{18})$ if all modes in $\mathcal{S}_R$ are excited when $R=1000$). However, this is still not sufficient due to the large number of elements in $Q$, with $|Q|\sim 10^8$. In order to further reduce the computational cost, we consider a small subset $Q_s$ of $Q$ (with $|Q_s|\sim 10^4$) that only includes the scale resonances, i.e., resonances generating wave modes with new length that are of vital importance to energy cascade. The remaining set $Q_a$ (with $|Q_a|\sim 10^8$) only contains angle resonances with pairwise equal lengths ($|\bm{k}_1|=|\bm{k}_3|$ and $|\bm{k}_2|=|\bm{k}_4|$; or $|\bm{k}_1|=|\bm{k}_4|$ and $|\bm{k}_2|=|\bm{k}_3|$) that are less important (but may play a role in connecting different scale resonances). Accordingly, we can formulate an efficient computational scheme to loop over $Q_s$ in each iteration and account for the resonances in $Q_a$ by its two bounds: an lower bound that no angle resonances are included and an upper bound that all modes with equal length with any active one are excited. Using this scheme, the computational cost for $R\sim 10^3$ becomes affordable and we are certain that the true solution lies between the two bounds of the computation, which is sufficient for the physical purpose in this study.

In this paper, we apply the computational method outlined above to study the resonant energy transfer of surface gravity waves in a finite discrete wavenumber space with $R=2500$. The initially active modes are placed in a circular region with radius $r_D$ for direct cascade. We show that for small $r_D$, a state of frozen turbulence is observed with the maximum reachable wavenumber $k_{max}$ linearly proportional to $r_D$. At $r_D \sim [60,110]$, a sharp transition to unlimited energy cascade occurs where $k_{max}\sim R$ is established for both upper and lower bounds. In addition to the direct cascade, we also study the inverse cascade with initial active modes in a ring-shaped area, and we show the transition from frozen turbulence to unlimited energy cascade (under their corresponding definitions in the context of inverse cascade) with the increase of the thickness of the ring. Finally, we analyze the structure of resonant quartets, through which the sharp transition from frozen turbulence to unlimited energy cascade is explained, and the role of angle resonances in determining the extent of the cascades is elucidated for both direct and inverse cascades.

\section{\label{sec:method}Methodology}
We consider the surface gravity wave system with resonant conditions in the form
\begin{equation}
    \begin{aligned}
    \bm{k}_1+\bm{k}_2&=\bm{k}_3+\bm{k}_4, \\
    \sqrt{k_1}+\sqrt{k_2}&=\sqrt{k_3}+\sqrt{k_4},
    \end{aligned}
\label{eq:gravityResonance}
\end{equation}
where $\bm{k}_i=(m_i,n_i)$ and $k_i=|\bm{k}_i|$ ($i=1,2,3,4$) with $m_i,n_i\in \mathbb{Z}$ and $\sqrt{m_i^2+n_i^2}\leq R$ for a given $R\in\mathbb{N}$. The solutions of \eqref{eq:gravityResonance} correspond to the exact resonances in a finite wavenumber domain $\mathcal{S}_R=\{\bm{k}\in\mathbb{Z}^2: k\leq R\}$. In our kinematic model, we specify the initial condition by defining a set of active modes denoted by $\mathcal{S}_0$. For the direct cascade, the initially active modes are placed in a circular area $\mathcal{S}_0=\{\bm{k}\in\mathbb{Z}^2: k\leq r_D\}$ where $0<r_D< R$. For the inverse cascade, they are placed in a ring-shaped area $\mathcal{S}_0=\{\bm{k}\in\mathbb{Z}^2: r_I\leq k\leq R\}$ where $0<r_I<R$. 

The solution for the direct/inverse cascade problem now amounts to iteratively finding the solution of \eqref{eq:gravityResonance} for energy spreading, i.e., for each iteration, we loop over all combinations of 3 active modes and check whether there exists a 4th mode satisfying \eqref{eq:gravityResonance}. Such a brute-force method, as reviewed, is not computationally affordable for large $R\sim 10^3$. Here we propose a novel and efficient method for this computation providing the upper and lower bounds of the true solution that are sufficient for our physical purpose. The first step is to pre-compute (before the start of the iterations) the set of scale resonances $\mathcal{Q}_s$ inside the spectral domain $\mathcal{S}_R$:
\begin{equation}
    \begin{aligned}
        \mathcal{Q}_s=\{(\bm{k}_1,\bm{k}_2,\bm{k}_3,\bm{k}_4):\bm{k}_1+\bm{k}_2=\bm{k}_3+\bm{k}_4,\\ \sqrt{k_1}+\sqrt{k_2}=\sqrt{k_3}+\sqrt{k_4},\\ 
        k_1\neq k_3,\ k_1\neq k_4,\\ 
        \bm{k}_i\in\mathcal{S}_R,\ \forall\ i=1,2,3,4 \}.
    \end{aligned}
\label{eq:scaleResonance}
\end{equation}
As shown by Kartashova, the scale resonances in $\mathcal{Q}_s$ are substantially sparser compare to all resonances (or angle resonances) \cite{kartashova2007exact,kartashova2008resonant}, and can be computed by a fast algorithm \cite{kartashova2006laminated}. The principle of this algorithm is that only a small subset of numbers need to be considered for frequency $\omega$ according to the irrationality of the dispersion relation, which substantially reduces the number of searches (since only the small subset and the corresponding wavenumbers need to be searched instead of all wavenumbers). We present the detailed algorithm to compute $\mathcal{Q}_s$ in the Supplemental Material \footnote{See Supplemental Material at [ ] for the computation of searching for all scale-resonant quartets in a finite discrete wavenumber space.} mainly following \cite{kartashova2006laminated}. The overall computational complexity can be estimated as $O(R^2)$ that only needs to be performed once for all kinematic calculations in this paper. 

With $\mathcal{Q}_s$ available, we can iteratively update the set of active modes $\mathcal{S}_i$ ($i=1,2,...,N$) starting from $\mathcal{S}_0$ with operation $\mathcal{S}_{i+1}=\boldsymbol{P}^s(\mathcal{S}_i)$. In particular, in performing $\boldsymbol{P}^s$, we loop over all the quartets in $\mathcal{Q}_s$, and if there exist three (and only three) active modes (according to $\mathcal{S}_i$) in one quartet, we activate the 4th one. The set $\mathcal{S}_{i+1}$ is then taken as the union of $\mathcal{S}_i$ and all newly activated modes. Since $\mathcal{Q}_s$ is invariant and sparse, the number of calculations at each iteration is a small constant, as opposed to the situation in the brute-force approach. 

To further account for the effects of angle resonances, two bounds of the result can be considered. The lower bound can be simply taken as $\mathcal{S}^l_i=\mathcal{S}_i$ ($i=1,2,...,N^l$), i.e., result of spreading only from scale resonances. For the upper bound, we consider the situation that at each iteration, after the spreading by scale resonances, we activate all modes in $S_R$ that have the same length with any active modes (defined as operation $\boldsymbol{P}^a$). More precisely, we perform $\mathcal{S}^u_{i+1}=\boldsymbol{P}^s(\mathcal{S}^u_i) \cup \boldsymbol{P}^a[\boldsymbol{P}^s(\mathcal{S}^u_i)]$ ($i=1,2,...,N^u$) from $\mathcal{S}^u_0=\mathcal{S}_0$, where the operation $\boldsymbol{P}^a$ is defined as $\boldsymbol{P}^a[\mathcal{S}^*]=\{\bm{k}: k=k_1,\forall \bm{k}_1\in\mathcal{S}^*\}$. The full algorithm for the computation is presented in algorithm \ref{alg:model}, with the code available on Github \footnote{https://github.com/joezhang13/ExactResonance}. 

\begin{algorithm}[H]
    \renewcommand{\algorithmicrequire}{\textbf{Input:}}
	\renewcommand{\algorithmicensure}{\textbf{Output:}}
	\caption{Kinematic model of energy cascades} 
	\label{alg:model} 
	\begin{algorithmic}
		\REQUIRE size of domain $R$, the radius $r_D$ for direct cascades or the inner radius $r_I$ for inverse cascades
		\ENSURE the final sets of active modes $\mathcal{S}^l_{N^l}$ for lower bound and $\mathcal{S}^u_{N^u}$ for upper bound
		\STATE \textbf{Initialization}  initialize the set of active modes $\mathcal{S}_0=\{\bm{k}\in\mathbb{Z}^2: k\leq r_D\}$ for direct cascades or $\mathcal{S}_0=\{\bm{k}\in\mathbb{Z}^2: r_I\leq k\leq R\}$ for inverse cascades
		\STATE \textit{Calculate $\mathcal{Q}_s$ from \eqref{eq:gravityResonance}}
		\STATE \textit{$i\gets 0$, $\mathcal{S}^l_0\gets \mathcal{S}_0$}
		\WHILE{$\mathcal{S}^l_i\neq \boldsymbol{P}^s(\mathcal{S}^l_i)$}
		\STATE $\mathcal{S}^l_{i+1}\gets \boldsymbol{P}^s(\mathcal{S}^l_i)$
		\STATE $i\gets i+1$
		\ENDWHILE
		\STATE $N^l\gets i$
		\STATE \textit{$i\gets 0$, $\mathcal{S}^u_0\gets \mathcal{S}_0$}
		\WHILE{$\mathcal{S}^u_i\neq \boldsymbol{P}^s(\mathcal{S}^u_i)$}
		\STATE $\mathcal{S}^u_{i+1}\gets \boldsymbol{P}^s(\mathcal{S}^u_i) \cup \boldsymbol{P}^a[\boldsymbol{P}^s(\mathcal{S}^u_i)]$
		\STATE $i\gets i+1$
		\ENDWHILE
		\STATE $N^u\gets i$
	\end{algorithmic}
\end{algorithm}

\section{\label{sec:results}Results}
\begin{figure}[htbp]
\includegraphics[width=0.45\textwidth]{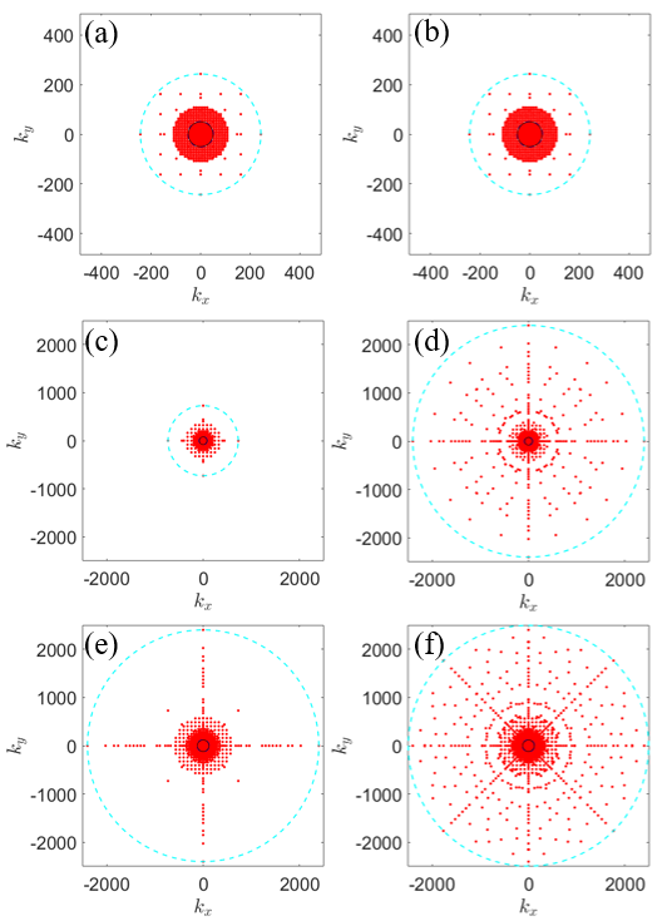}
\caption{The distribution of active wave modes (red dots) in the final state of direct cascades in a domain with $R=2500$, for (a) $r_D=50$, lower bound, (b) $r_D=50$, upper bound, (c) $r_D=80$, lower bound, (d) $r_D=80$, upper bound, (e) $r_D=120$, lower bound, and (f) $r_D=120$, upper bound. The initially excited region is indicated by a red area circled by a black solid line. The maximum wavenumber $k_{max}$ of the active modes is indicated by a circle with cyan dashed line.}
\label{fig:direct}
\end{figure}

We first present results for direct cascade in a wavenumber domain of $R=2500$ using our kinematic model. Figure \ref{fig:direct} shows the upper and lower bounds of final distributions of active modes (i.e., no more activated modes in the next iteration) for three cases with $r_D$=$50,80,120$. For $r_D=50$ as in (a) and (b), the results from the upper and lower bounds are identical, indicating a true solution of frozen turbulence with maximum reachable wavenumber $k_{max} = 243$. For $r_D=80$ as in (c) and (d), difference in the upper and lower bounds of the solution is observed, showing that the true solution $k_{max}$ lies in a relatively large range of $[729,2401]$, still a state of frozen turbulence according to our definition. Finally, for $r_D=120$ as in (e) and (f), the upper and lower bounds become much closer, both close to the outer boundary of the domain with $k_{max}\sim R$, indicating a state of unlimited energy cascade. This is in spite of the fact that the upper-bound solution is associated with a much denser active modes excited by angle resonances. 

\begin{figure}[htbp]
\includegraphics[width=0.4\textwidth]{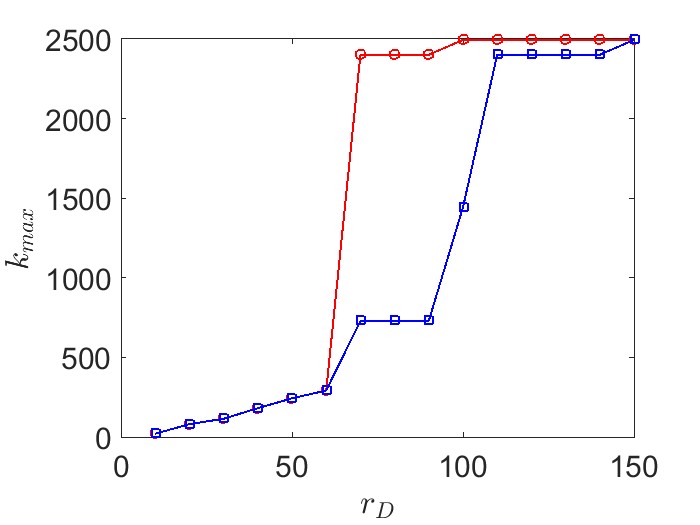}
\caption{The maximum wavenumber $k_{max}$ as a function of the radius of the initial region $r_D$ in direct cascades with $R=2500$. Both upper (red line with squares) and lower (blue line with squares) bounds are shown.}
\label{fig:kmax}
\end{figure}

A more complete view of the relation between $k_{max}$ and $r_D$ is shown in figure \ref{fig:kmax} with both upper and lower bounds of the solution. We see that there exists a sharp transition between frozen turbulence and unlimited energy cascade for a critical value of $r_D \sim [60,110]$ (considering the uncertainty exhibited between upper and lower bounds). For $r_D$ below the critical value, a frozen turbulence state is observed with exact match of upper and lower bounds of the solution, showing a linear relation fitted by $k_{max}\approx 5.5r_D-35.9$. For $r_D$ above the critical value, the state quickly transits to unlimited energy cascade with $k_{max} \gg r_D$ (and eventually $k_{max}\sim R$). Such a transition implies a bifurcation of solutions depending on the initial condition of energy distribution in discrete turbulence of surface gravity waves. In addition, we remark that the frozen turbulence observed here for small $r_D$ is consistent with the simulation of primitive Euler equations at low nonlinearity level \cite{zhang2022numerical}, which suggests a frozen turbulence state when the forcing is located in the range of $k\in[1,19]$. 

\begin{figure}[htbp]
\includegraphics[width=0.45\textwidth]{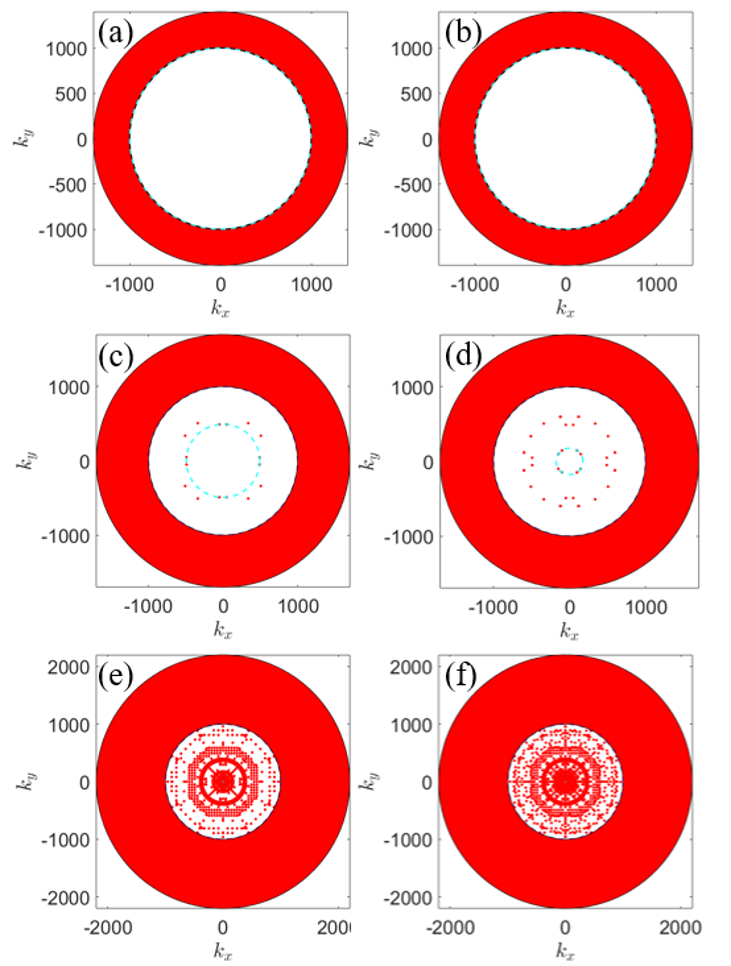}
\caption{The distribution of active wave modes (red dots) in the final state of inverse cascades with $r_I=1000$, for (a) $R=1400$, lower bound, (b) $R=1400$, upper bound, (c) $R=1700$, lower bound, (d) $R=1700$, upper bound, (e) $R=2200$, lower bound, and (f) $R=2200$, upper bound. The initially excited region is indicated by the ring-shaped red area. The minimum wavenumber $k_{min}$ of the active modes is indicated by a circle with cyan dashed line.}
\label{fig:inverse}
\end{figure}

We next present the results for inverse cascade with initial active modes placed in a ring with radius $[r_I, R]$. Our interest is to understand the effect of ring thickness on the minimum wavenumber $k_{min}$ reachable in the cascade. For this purpose, we keep $r_I=1000$ as a constant and vary $R$ to examine its effect on the inverse energy cascade. Figure \ref{fig:inverse} shows the upper and lower bounds of final distributions of active modes in inverse cascades for $R$=$1400,1700$ and $2200$. For $R=1400$ as in (a) and (b), a frozen turbulence state is observed with no new active modes generated (for both upper and lower bounds). For $R=1700$ as in (c) and (d), $k_{min}$ reaches $[492.4,176.7]$ according to the two bounds of solution, indicating again a state of frozen turbulence. The unlimited energy cascade is enabled as in (e) and (f) for $R=2200$, where $k_{min}\sim 1$ (the fundamental mode) from both the upper and lower bounds. 

\begin{figure}[htbp]
\includegraphics[width=0.4\textwidth]{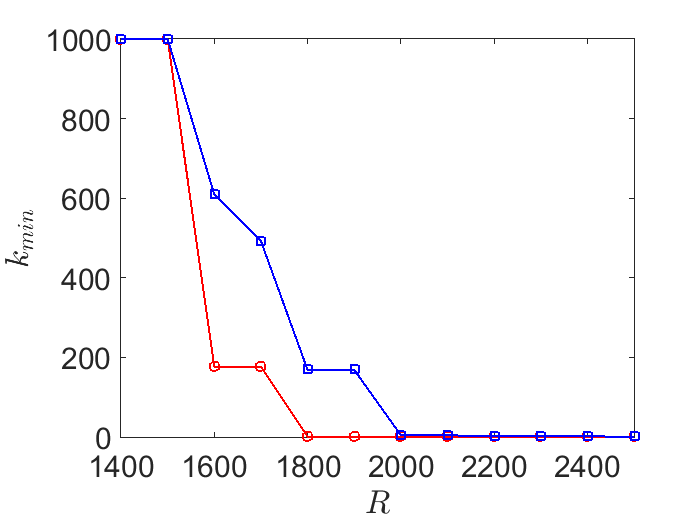}
\caption{The minimum wavenumber $k_{min}$ as a function of the outer radius $R$ in inverse cascades with $r_I=1000$. Both upper (red line with squares) and lower (blue line with squares) bounds are shown.}
\label{fig:kmin}
\end{figure}

We show a more complete view of the relation between $k_{min}$ and $R$ in figure \ref{fig:kmin}. For $R\leq 1500$, no new modes can be generated. As $R$ increases beyond 1500, discrepancy between upper and lower bounds occurs, with step-wise transition behavior shown in the solutions. The frozen turbulence state transits to unlimited energy cascade for $R \in [1800, 2000]$, after which the upper and lower bounds become close to each other with $k_{min}$ close to unity.

Finally, by examining the behavior of the upper and lower bounds in direct and inverse cascades, the role of angle resonances in the cascades can be inferred. It is clear in both cases that discrepancy between the two bounds occurs mainly when the initially active region is of appropriate size (i.e., not too small or too large). This is partly expected due to the finite wavenumber space, i.e., if the scale-resonance cascade can reach the boundary, angle resonances are not needed (so that the upper and lower bounds for large initial regions are guaranteed to be the same). However, it will be beneficial to conduct a more rigorous analysis to elucidate the full mechanism, as well as the sharp transition from frozen turbulence to unlimited energy cascade above $r_D\approx 60$ for direct cascade. Such an analysis is presented in \S IV.

\section{\label{sec:discussion}Structure of resonant quartets}
In order to understand the state transition and role of angle resonances, it is necessary to discuss the structure of scale resonances in $\mathcal{Q}_s$. In particular, we aim to quantify the capability of a given mode in connecting smaller and larger scales through scale resonances, which also provides hint on the role of angle resonances at this scale.

For this purpose, we define the scale-resonance multiplicity index $M_s(\bm{k})$ and order index $O_s(\bm{k})$
\begin{equation}
    M_s(\bm{k})=|\{q:\bm{k}\in q\}|, \ \ O_s(\bm{k})= \overline{\{RK[\bm{k};q]:\bm{k}\in q\}}
\label{eq:Ms}
\end{equation}
where $q=\{\bm{k}_1,\bm{k}_2,\bm{k}_3,\bm{k}_4\}$ is a set taken from the quartets in $\mathcal{Q}_s$, $|\cdot|$ is the cardinality of the set, which in this context counts the number of $q$ for which $\bm{k}$ is an element, $RK[\bm{k};q]$ gives the ranking index of $\bm{k}$ in set $q$ (1,2,3,4 for length from smallest to largest), and $\overline{\ \cdot\ }$ calculates the average of the numbers in the set, and in this context the average of ranking indices. With definition \eqref{eq:Ms}, $M_s(\bm{k})$ measures the number of scale resonances including mode $\bm{k}$ and $O_s(\bm{k})$ measures the average ranking of $\bm{k}$ (in terms of length) among all scale-resonant quartets. For a favorable situation of energy cascade crossing $\bm{k}$, it is desired that $M_s(\bm{k})$ is large and $O_s(\bm{k})$ is in the middle range (say around 2.5).  

\begin{figure*}[htbp]
\includegraphics[width=0.9\textwidth]{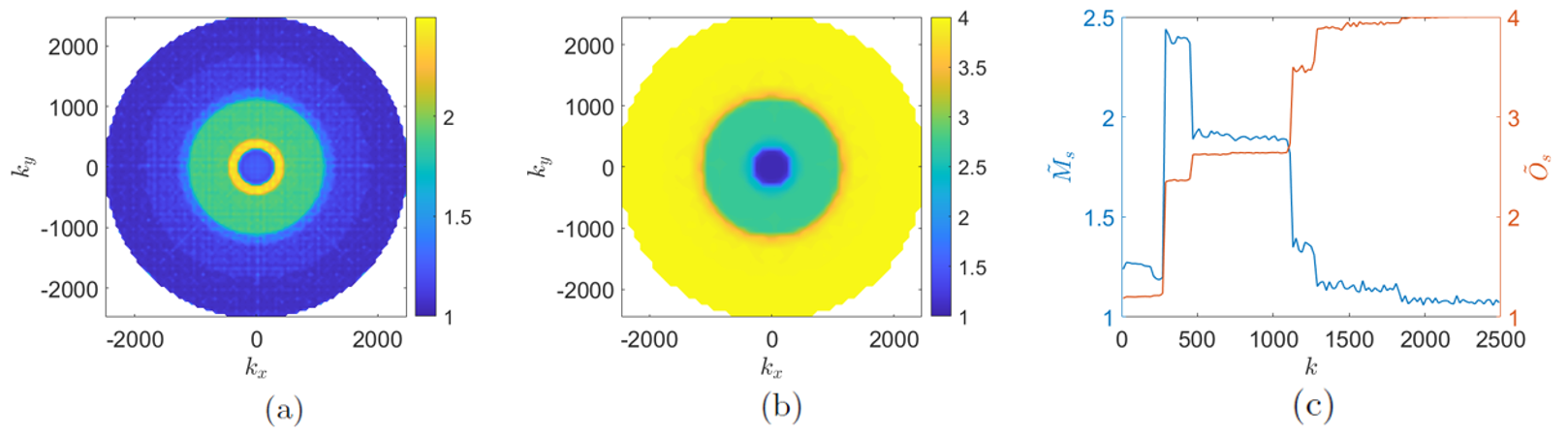}
\caption{The distribution of (a) $M_s$ and (b) $O_s$ in the $(k_x,k_y)$ plane, and (c) $\tilde{M}_s$ and $\tilde{O}_s$ as functions of $k$ with $R=2500$. In (a) and (b), the values are averaged over the modes with nonzero $M_s(\bm{k})$ in a $50\times 50$-grid area to avoid large discrete peaks in the original field.}
\label{fig:Ms}
\end{figure*}

Figure \ref{fig:Ms}(a) and (b) show $M_s(\bm{k})$ and $O_s(\bm{k})$ in the domain $\mathcal{S}_{R=2500}$. We note that in the plot we only consider those modes for which $M_s(\bm{k})$ is nonzero, i.e., there exist some quartets in $\mathcal{Q}_s$ for which $\bm{k}$ is an element. In general, it can be seen that the function $M_s(\bm{k})$ peaks around the middle range of $k$,  whereas $O_s(\bm{k})$ increases from 1 to 4 as $k$ increases. The behaviors of $M_s(\bm{k})$ and $O_s(\bm{k})$ at large $k$ can be partly explained by the finite wavenumber space $\mathcal{S}_R$, which eliminates many quartets (with modes beyond $R$), leading to $M_s(\bm{k})\approx 0$ and $O_s(\bm{k})\approx 4$. In order to more precisely quantify the two indexes at each (scalar) scale $k$, we further define the angle-averaged quantities:

\begin{equation}
    \tilde{M}_s(k)=\overline{\{M_s(\bm{k}^*):||\bm{k^*}|-k|<\delta k/2,M_s(\bm{k}^*)\neq 0\}},
\end{equation}
\begin{equation}
    \tilde{O}_s(k)=\overline{\{O_s(\bm{k}^*):||\bm{k^*}|-k|<\delta k/2,M_s(\bm{k}^*)\neq 0\}},
\end{equation}
with $\delta k=20$ used in our calculation. We plot $\tilde{M}_s(k)$ and $\tilde{O}_s(k)$ in figure \ref{fig:Ms}(c), which shows consistent behavior as (a), (b), and helps elucidate the transition from frozen turbulence to unlimited energy cascade, as well as the role of angle resonances, as explained below.

In both direct and inverse cascades, if the initial active region is small (i.e., $r_D$ and $R$ is small for respectively the direct and inverse cascades), the energy cascade dies out close to the initial region. This is because the index $\tilde{O}_s(k)$ is either close to 1 and 4 for direct and inverse cascade, and the index $\tilde{M}_s(k)$ is small for both cases, resulting in an unfavorable situation of generating new scales (by scale resonances). Under this situation, even if many angle resonances are included (in fact for initial state the modes are already dense in the angular direction), the cascade will not be further excited. As the initial active region grows to appropriate size, $k_{max}$ and $k_{min}$ can reach the middle range in $\mathcal{S}_R$, where the scale of favorable cascade becomes active, i.e., scales with $\tilde{O}_s(k)\approx 2.5$ and large $\tilde{M}_s(k)$. We see from \ref{fig:Ms}(c) that this favorable scale of energy transfer occurs as $k$ grows above 250. Referring to figure \ref{fig:kmax}, $k=250$ corresponds to $r_D\approx 60$, which is exactly the starting point of the sharp transition from frozen turbulence to unlimited energy cascade. For these cases, angle resonances become important as they excite more modes in the favorable scale $k$ to sustain the cascade, leading to discrepancies in the upper and lower bounds of the solution. Finally, as the initial active region becomes large enough so that scale resonances 
themselves can sustain an energy cascade to the boundary of the domain (i.e., $k_{max}$ reaching $R$ and $k_{min}$ reaching 1), angle resonances become not important for the extent of cascades. More precisely, they result in difference in the density of active modes but do not affect $k_{max}$ and $k_{min}$.

\section{\label{sec:conclusion}Conclusions}
In this paper, we study the energy cascade of surface gravity waves by exact resonances in a finite discrete wavenumber space. A kinematic model is developed based on the scale resonances and the upper and lower bounds of angle resonances, which significantly reduces the computational cost compared to the brute-force searching method used in previous studies. Both direct and inverse cascades are examined with different sizes of initially active regions. For direct cascade, we find that there exists a critical radius of the initial region within the range $[60,110]$, above which the dynamics transits from frozen turbulence to unlimited energy cascade (with cascade extending to the boundary $R=2500$). For inverse cascade, states of frozen turbulence and unlimited energy cascade are also observed but with a somewhat less sharp transition. We finally analyze the structure of resonant quartets, which helps to elucidate the sharp transition in $[60,110]$ and the role of angle resonances in the extent of the cascades observed in the kinematic studies.



\bibliography{references}

\end{document}


\maketitle 

In this document, we provide detailed description of the algorithm used in this study to compute all the scale-resonant quartets in a finite discrete wavenumber domain. 

To find all the scale resonances satisfying the constraints of resonant interactions, we follow a generic method proposed by \cite{kartashova2006laminated} to the system of surface gravity waves with the resonant conditions taking the form 
\begin{equation}
    \begin{aligned}
    \bm{k}_1+\bm{k}_2&=\bm{k}_3+\bm{k}_4, \\
    \sqrt{k_1}+\sqrt{k_2}&=\sqrt{k_3}+\sqrt{k_4},
    \end{aligned}
\label{eq:gravityResonance}
\end{equation}
where $\bm{k}_i=(m_i,n_i)$ and $k_i=|\bm{k}_i|$ ($i=1,2,3,4$) with $m_i,n_i\in \mathbb{Z}$ and $\sqrt{m_i^2+n_i^2}\leq R$ for a given $R\in\mathbb{N}$. The main idea of this method is to partition the spectral domain into disjoint classes of $\bm{k}$ which allows us to search for solutions in each class efficiently.

We first rewrite \eqref{eq:gravityResonance} in the following form:
\begin{equation}
    \begin{aligned}
        m_1+m_2&=m_3+m_4, \\
        n_1+n_2&=n_3+n_4, \\
        \omega_1+\omega_2&=\omega_3+\omega_4,
    \end{aligned}
\label{eq:resonanceScalar}
\end{equation}
where $\omega_i=|\bm{k}_i|^{1/2}=(m_i^2+n_i^2)^{1/4}$ ($i=1,2,3,4$).

Now let's consider a set of algebraic numbers $\omega=t^{1/4}$, $t\in\mathbb{N}$. Any such number $\omega$ can be represented by
\begin{equation}
    \omega=\gamma q^{1/4},\quad \gamma\in\mathbb{N},
\label{eq:wq}
\end{equation}
where $q$ is a product
\begin{equation}
    q=p_1^{e_1}p_2^{e_2}...p_n^{e_n},
\label{eq:qp}
\end{equation}
with $p_1,p_2,...,p_n$ different primes and the powers $e_1,e_2,...,e_n\in\mathbb{N}$ all smaller than 4. Then we define the set of numbers $\omega$ with the same $q$ as the $q$-class denoted by $Cl_q$, where $q$ is called the class index. For each $\omega=\gamma q^{1/4}\in Cl_q$, $\gamma$ is called the weight of $\omega$. 

Based on these definitions, it can be shown that there are two types of solutions for \eqref{eq:resonanceScalar}:

\textbf{Case 1.} All the numbers $\omega_i$ ($i=1,2,3,4$) belong to the same class $Cl_q$. In this case the third equation of \eqref{eq:resonanceScalar} can be written as
\begin{equation}
    \gamma_1 q^{1/4}+\gamma_2 q^{1/4}=\gamma_3 q^{1/4}+\gamma_4 q^{1/4}
\label{eq:case1}
\end{equation}
with $\gamma_i\in\mathbb{N}$ ($i=1,2,3,4$).

\textbf{Case 2.} The numbers $\omega_i$ belong to two different classes $Cl_{q_1}$ and $Cl_{q_2}$. In this case the third equation of \eqref{eq:resonanceScalar} can be written as
\begin{equation}
    \gamma_1 q_1^{1/4}+\gamma_2 q_2^{1/4}=\gamma_1 q_1^{1/4}+\gamma_2 q_2^{1/4}
\label{eq:case2}
\end{equation}
with $\gamma_i\in\mathbb{N}$ ($i=1,2$).

It is evident that scale resonances, which generate modes with new lengths, are only possible in Case 1 since Case 2 consists of modes with pairwise equal lengths. Therefore, we concentrate on finding possible solutions in Case 1. The general idea is to take all solutions of $\gamma_1+\gamma_2=\gamma_3+\gamma_4$ with $\gamma_i^4 q$ decomposable into the sum of two squares $\gamma_i^4 q=m_i^2+n_i^2$ and then check the linear conditions (the first and second equations in \eqref{eq:resonanceScalar}). The detailed description of this algorithm is as follows.

\section{Procedures}
\subsection{Calculating class indexes}
We first consider numbers $t_i=\omega_i^4=\gamma_i^4 q$. To obtain a solution for \eqref{eq:resonanceScalar}, $t_i$ must have a representation as the sum of two squares of integers, i.e., $t_i=m_i^2+n_i^2$. This requirement restricts the class index $q$ in the following way: according to the Euler's theorem \citep{euler1763Theoremata}, an integer can be represented by the sum of two squares if and only if every factor with the form $p\equiv 4u+3$ ($u\in\mathbb{N}$) contained in its prime factorization is in an even degree. As $\gamma_i^4$ already contains every prime factor in an even degree (the degree of any factor is an integer multiple of 4), this condition must be held for $q$. Note that by construction, all prime factors of $q$ are in degrees smaller than 4. Therefore, the restriction for $q$ becomes: if $q$ is divisible by a prime $p\equiv 4u+3$, it should be divisible by its square and not be divisible by its cube.  

Based on the above requirements, the possible class indexes $q$ for modes inside the spectral domain $\mathcal{S}_R$ ($q\leq R^2$) are calculated as follows. We create a list of integers $A_q=\{1,2,...,R^2\}$ and make 2 passes over it. In the first pass, we remove all the numbers divisible by the 4th power of any prime. In the second pass, we examine the numbers divisible by any prime satisfying $p\equiv 3\mod 4$: if the number is not divisible by $p^2$ or divisible by $p^3$, we remove it from $A_q$. Finally, $A_q$ contains all the class indexes $q$ that can be represented by the sum of two squares.

\subsection{Solving the equation for weights}
The next step is to solve the equations for the weights obtained from \eqref{eq:case1}
\begin{equation}
    \gamma_1+\gamma_2=\gamma_3+\gamma_4
\label{eq:gamma}
\end{equation}
with $1\leq \gamma_i\leq M(q)$, where $M(q)=\lfloor (R^2/q)^{1/4}\rfloor$ is the maximum possible weight $\gamma$ for a given $q$ with $\lfloor\cdot\rfloor$ being the round-down operator. Let $S_\gamma=\gamma_1+\gamma_2=\gamma_3+\gamma_4$. Without loss of generality we can suppose
\begin{equation}
    \gamma_1\leq\gamma_3\leq\gamma_4\leq\gamma_2.
\end{equation}
There are 4 possible cases for the weights: 

1. $\gamma_1<\gamma_3<\gamma_4<\gamma_2$, 

2. $\gamma_1=\gamma_3<\gamma_4=\gamma_2$,

3. $\gamma_1<\gamma_3=\gamma_4<\gamma_2$,

4. $\gamma_1=\gamma_3=\gamma_4=\gamma_2$.

\noindent Since our aim is to find the scale resonances, we only need to consider cases 1 and 3 where the lengths of the four modes are not pairwise equal. Then the search of the solutions of \eqref{eq:gamma} is basically to find 2 partitions of $S_\gamma$ ($\gamma_1+\gamma_2$ and $\gamma_3+\gamma_4$) satisfying cases 1 and 3, which can be computed straightforwardly by looping over all possible combinations of the partitions of $S_\gamma$ under the condition that $1\leq\gamma_i\leq M(q)$ ($i=1,2,3,4$).

Note that before this step, we can reduce the computational cost by discarding classes which consist of no solution in cases 1 and 3. Consider the classes with $M(q)=1$, this means that all 4 weights $\gamma_i=1$ ($i=1,2,3,4$). These classes correspond to the case 4, which have all 4 modes with the same length. Therefore, we can directly discard such classes without any computation of solving \eqref{eq:gamma}. Similarly, we can also discard classes with $M(q)=2$ since only cases 2 and 4 are possible in this condition. According to our calculation, there are $98.5\%$ class with $M(q)=1,2$ among all the classes obtained in the domain $\mathcal{S}_{R=1000}$. Thus, discarding them significantly reduces the number of classes to be checked in the following steps.

\subsection{Decomposition into sum of squares}
The aim of this step is to find the decompositions of the number $\gamma_i^4 q$ with $\gamma_i=1,2,...,M(q)$. The generalized form of this problem has been investigated by \cite{basilla2004solution}, who provides an efficient algorithm to search for all decompositions of $b=x^2+y^2$ ($x,y\in\mathbb{N}$). First, we search for all $a\in\mathbb{N}$ satisfying $a^2\equiv (-1)\mod b$, $0<a<b/2$. Then for each $a$ we construct a finite sequence $\{r_j\}$ with $r_0=b$, $r_1=a$, $r_{j+2}=\beta_j r_{j+1}-r_j$, and $\beta_j=\lfloor r_j/r_{j+1}\rfloor$. It can be shown that $\exists N>1$ ($N\in\mathbb{N}$) such that $r_0>r_1>...>r_N=1>r_{N+1}=0$. If we can find an integer $k$ such that $r_{k-1}^2>b>r_k^2$, then $b=r_k^2+r_{k+1}^2$. Based on this method, we can obtain all the decomposition of $\gamma_i^4 q$ into sum of two squares of integers. Note that if $b=c^2(x^2+y^2)$ with $c>1$ ($c\in\mathbb{N}$), the above method cannot obtain such decomposition directly. Therefore, we need to check if $\gamma_i^4 q$ is divisible by an integer square before applying this method. In such conditions, we divide $b$ by $c^2$ first and then compute the decomposition using the above method.

\subsection{Checking linear conditions}
Finally, we check the linear conditions (the first and second equations in \eqref{eq:resonanceScalar}) to find all solutions based on $(m_i^*,n_i^*)$ ($m_i^*\geq 0$, $n_i^*\geq 0$, $i=1,2,3,4$) satisfying the third equation of \eqref{eq:resonanceScalar} obtained from the previous steps. This is done by taking all combinations of signs satisfying
\begin{equation}
\begin{aligned}
    &\pm m_1^*\pm m_2^*=\pm m_3^*\pm m_4^*, \\
    &\pm n_1^*\pm n_2^*=\pm n_3^*\pm n_4^*.
\end{aligned}
\label{eq:linear}
\end{equation}
There are $2^8=256$ combinations in total for each set of $(m_i^*,n_i^*)$. With the repeated solutions (e.g., some $m_i^*$ or $n_i^*$ being 0) and the symmetric solutions (e.g., all signs become opposite in the equations) taken into consideration, a correct, exhaustive and efficient search can be constructed.

\section{Summary}
The full computation process is summarized in algorithm \ref{alg:solver}.

\begin{algorithm} 
    \renewcommand{\algorithmicrequire}{\textbf{Input:}}
	\renewcommand{\algorithmicensure}{\textbf{Output:}}
	\caption{Search for scale resonances} 
	\label{alg:solver} 
	\begin{algorithmic}
		\REQUIRE size of the domain $R$
		\ENSURE set of scale resonance quartets $\mathcal{Q}_s$
		\STATE 1. Compute the list of class indexes $A_q=\{q:q\in\mathbb{N}^+,q\leq R^2;\ \exists\ x,y\in\mathbb{N}^+, q=x^2+y^2\}$.
		\STATE 2. Solve \eqref{eq:gamma} to obtain all possible $\gamma_i$ ($i=1,2,3,4$) for each $q\in A_q$.
		\STATE 3. Decompose $\gamma_i^4 q$ obtained in 1 and 2 into sum of two squares: $m_i^{*2}+n_i^{*2}$ ($m_i^{*},n_i^{*}\in\mathbb{N}$).
		\STATE 4. Check the linear conditions \eqref{eq:linear} and put all qualified quartets $(\bm{k}_1,\bm{k}_2,\bm{k}_3,\bm{k}_4)$ with $\bm{k}_i=(\pm m_i^*, \pm n_i^*)$ into $\mathcal{Q}_s$.
	\end{algorithmic} 
\end{algorithm}

\bibliographystyle{apalike}
\bibliography{references}